\documentclass[multphys,vecphys]{svmult}

\usepackage{makeidx}         
\usepackage{graphicx}        
\usepackage{multicol}        
\usepackage[bottom]{footmisc}

\makeindex             

\begin{document}

\title*{Formation of cold molecular filaments in cooling flow clusters}
\author{Yves Revaz\inst{1},
Fran\c coise Combes\inst{1}\and Philippe Salom\'e\inst{2}}
\institute{LERMA, Observatoire de Paris, 61 av. de l'Observatoire, 75014 Paris, France
\and IRAM, 300 rue de la piscine, 38400 St-Martin d'H\`eres, France}
%
%
\maketitle


\section{Introduction}

New CO observations at the center of cooling flow in Perseus cluster \cite{salome06} 
show a clear correlation of the molecular gas with the previously detected H-$\alpha$ filaments \cite{conselice01}. 
In this poster, we present high resolution multi-phase simulations of the Perseus Cluster, taking into account
the AGN feedback in form of hot buoyant bubbles. 
These simulations show that significant amount of gas can cool far from the center. The AGN feedback
provides some heating, but also trigger the hot gas compression, that favours cooling (positive feedback), 
even at high radius ($R>30\,\rm{kpc}$). 
The cooled gas flows into the cluster core forming the observed filaments.

\section{Initial Conditions and gas physics}

Our cluster model is designed to fit the Perseus X-ray data \cite{sanders04}. 
The total mass distribution profile follows a pseudo-isothermal sphere with a
central electronic density of $5\times 10^{-2}$ and a core radius of $40\,\textrm{kpc}$.
The mass distribution is truncated at $2\,\rm{Mpc}$ and the total mass is $5.5\times 10^{14}\rm{M_\odot}$. 
The gas corresponding to 15\% of the total mass has an initial temperature of  $2.8\times 10^7\,\rm{K}$. 
%

The bulk of the intra-cluster gas has at high temperature ($10^7\,\rm{K}$) is well 
modeled by an ideal gas with adiabatic index of $5/3$. This hot phase has been
computed using and SPH technique.
For lower temperatures, SPH is not suited to render the clumpiness of the gas. 
Below $10^4\,\rm{K}$, the gas is treated as semi-collisional, using a sticky particle 
technique \cite{combes85}. Star formation may occur in this cold phase, following a Schmidt law. 
The cooling of the hot gas is computed using the standard cooling function \cite{sutherland93}, 
assuming an abundance of one third of the solar one.
%

Following \cite{sijacki06}, the AGN feedback is modeled by injecting energy in 
hot intra-cluster gas bubbles which are then driven buoyantly at higher radius. 
Bubbles are defined by their position, diameter, temperature, over pressure
and angle. They are generated by symmetric pairs, in average each $200\,\rm{Myr}$. 
%
%
%
%
%
%
%

The simulations have been run using a modified version of the Gadget2 Tree-SPH code (Springel, 2005), 
including cooling, star formation and multiphase SPH/Sticky gas. 
In each simulations the gas is represented by 4'194'304 SPH particles. 
The dark matter is modeled by a fixed outer potential.

\section{Results}

\begin{figure}
\centering
\includegraphics[angle=-90]{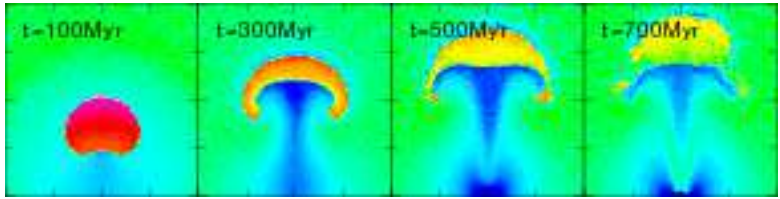}
\caption{Temperature map of an isolated bubble. From $10^4\,\rm{K}$ to $2\times 10^8\,\rm{K}$.
The box dimension is $200\times200\,\rm{kpc}$.}
\label{fig4}
\end{figure}

Fig.~\ref{fig4}. shows the evolution of an isolated bubble.
The bubble has an initial temperature of $10^8\,\rm{K}$.
It reaches $30\,\rm{kpc}$ and takes the characteristic mushroom shape as observed in H-$\alpha$ and 
$X$-ray \cite{fabian03}. 
A substantial amount of cooler gas 
(about $1/2$ of the bubble mass) 
is dragged by the rising bubble and forms 
the trunk of the mushroom. Between $t=300-500\,\rm{Myr}$, this gas falls back into the cluster center, 
while the head of the mushroom stops and dissipate into the intra-cluster medium. 

\vspace{-0.5cm}
\subsection{Global evolution and cold gas formation}
\vspace{-0.5cm}
\begin{figure}
\centering
\includegraphics[width=4.6cm,angle=-90]{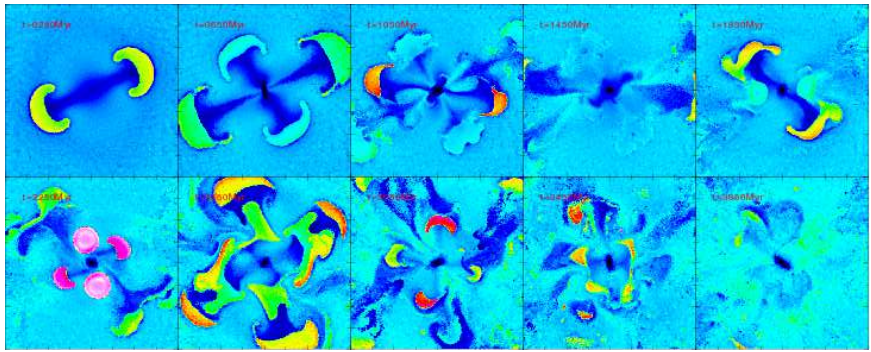}
\caption{Global evolution of  the cluster temperature during $4\,\rm{Gyr}$.
The box dimension is $400\times400\,\rm{kpc}$.
}
\label{fig1}
\end{figure}

\begin{figure}
\centering
\includegraphics[height=5.5cm]{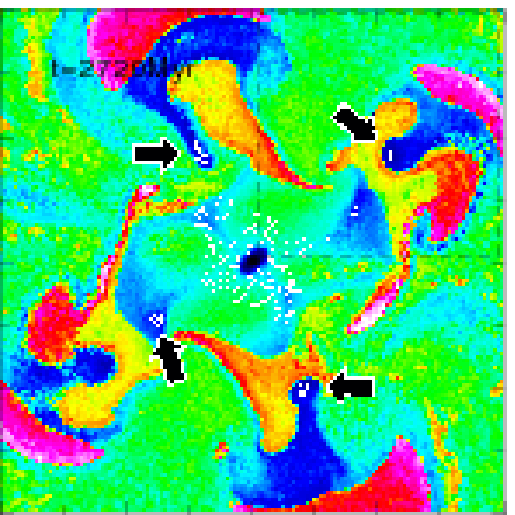}
\includegraphics[height=5.5cm]{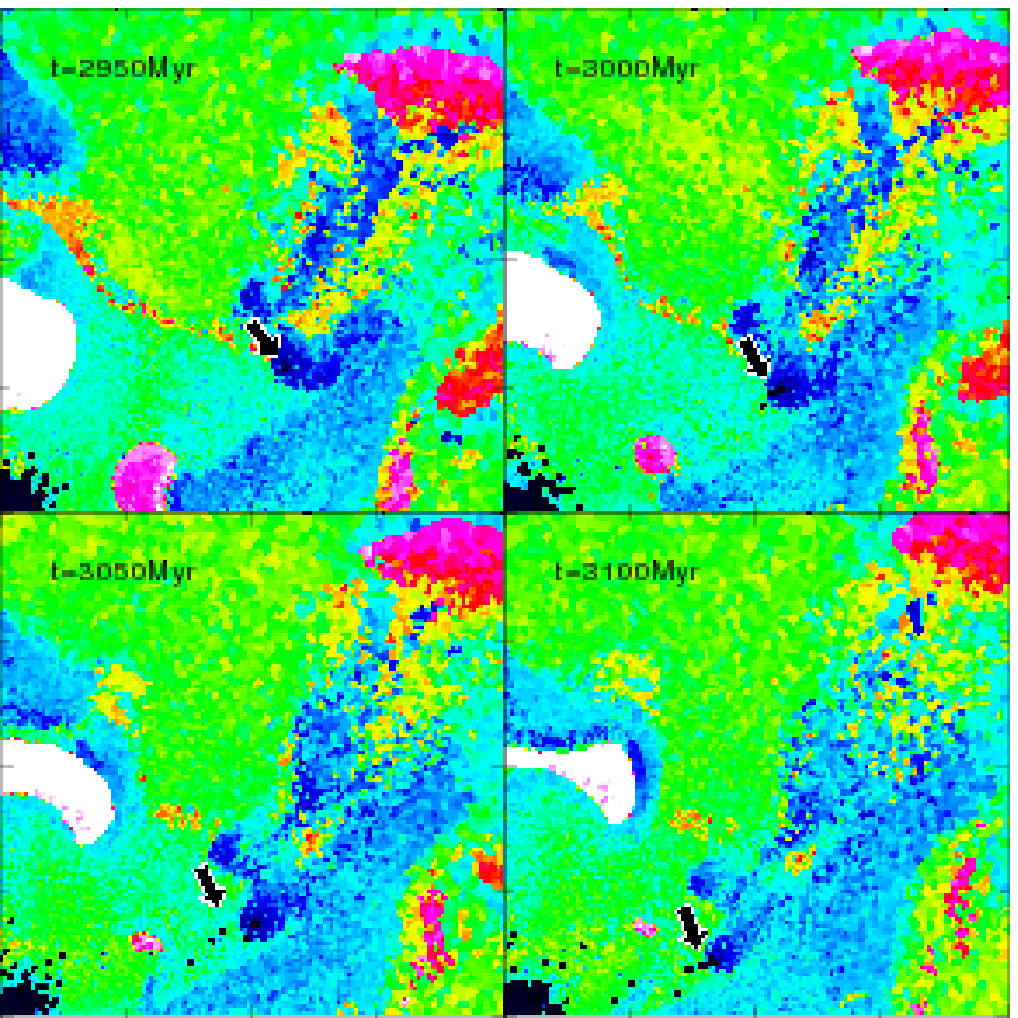}
\caption{
\emph{Left~:} Gas temperature at $t=2720\,\rm{Myr}$.
The white contours pointed by black arrows indicate the position of the gas having a short 
cooling time $\tau_{\rm{c}} <1.1\,\rm{Gyr}$ that will be transformed into cold gas.
The upper left box corresponds to the region zoomed in the right part of the figure.
\emph{Right~:} Gas temperature between $t=2950$ and $t=3100\,\rm{Myr}$,
from $10^4$ to $2\times 10^8\,\rm{K}$ .
The small black dots pointed by the black arrow represents cold gas ($T<10^4\,\rm{K}$)
falling into the cluster center.}
\label{fig3}
\end{figure}

In addition to slowing down the cooling flow at the center of the cluster, our simulation also show how AGN
feedback may trigger cold gas formation at high radius ($R>30\,\rm{kpc}$). Physical conditions for
efficiently cool intra-cluster gas occurs either when the gas is less than $10^7\,\rm{K}$ or when its density is sufficiently high. 
Buoyant bubbles are responsible of strong inhomogeneities in temperature (see Fig.~\ref{fig1}) as well as in density
of the intra-cluster medium. In Fig.~\ref{fig3} left, we show how cooler falling gas (trunk of a bubble, see Fig.~\ref{fig1}) is trapped 
between an old and a new rising bubble and is compressed to reach a state where its cooling time is sufficiently
short to let the gas becomes cool in a fraction of Gyr. In Fig.~\ref{fig3} right, the cluster is seen $300\,\rm{Myr}$ later.
The short cooling time gas of Fig.~\ref{fig3} left, has now cooled down below $10^4\,\rm{K}$. As it is not supported by the hot gas pressure,
it falls radially to the center, forming a filament like structure ($50$ to $100\,\rm{kpc}$) of mass $1.5\times 10^{8}\rm{M_\odot}$.
In the filament, the gas density is not high enough to form stars.  

\vspace{-0.2cm}

\vspace{-1cm}

%
%

%
%



\printindex
\end{document}